\documentstyle[12pt,a4,epsfig]{article}
\renewcommand{\theequation}{\thesection.\arabic{equation}}
\begin{document}
\begin{titlepage}
\begin{flushright}
DO-TH-95/18\\
November 1995
\end{flushright}

\vspace{20mm}
\begin{center}
{\Large \bf
Computing numerically the functional derivative of an effective action}

\vspace{10mm}

{\large  J. Baacke\footnote{
e-mail:~baacke@zx2.hrz.uni-dortmund.de} and A. S\"urig
\footnote{e-mail:~suerig@hal1.physik.uni-dortmund.de}} \\
\vspace{15mm}

{\large Institut f\"ur Physik, Universit\"at Dortmund} \\
{\large D - 44221 Dortmund , Germany}
\vspace{25mm}

\bf{Abstract}
\end{center}
The functional derivative of the effective action with respect to
an external field is part of the equation of motion of this field
if one-loop effects induced by
quantum fluctuations or thermal fluctuations are included in
minimizing the action of this field. Examples occur in all
field theories displaying classical solutions or - as the
Nambu-Jona-Lasinio model - selfconsistent field configurations.
We describe here a numerical method for computing such functional
derivatives; we use a fermion field with Yukawa interaction to an
external field as an example which is sufficiently simple and
sufficiently general. We compare the computed functional derivative
to analytical estimates.
\end{titlepage}


\section{Introduction}
\setcounter{equation}{0}
The one-loop effective action plays an important r\^ole in various
contexts of quantum field theory. It leads to a correction to
the energy of classical solutions
(see e.g. \cite{Raj}) and to corrections to quantum
tunneling \cite{CaCo} and thermal transition rates
[3-5]; it is the basis of
Hartree type selfconsistent field equations. If in the former case
the corrections are important the effective action should be
included in determining the extremal classical field configuration.
In Hartree type computations as e.g. in the nontopological
soliton model of Friedberg and Lee \cite{FrieLee,Baa79} or in the
Nambu-Jona-Lasinio model [8-11]
 the classical action may even have no
solution and the classical field may have no genuine action, it may be
just a parametrization of condensates.
Denoting the ``classical'' field as $\phi$ the equation of motion
reads
\begin{equation} \label{eqmot}
\frac{\delta \Gamma}{\delta \phi({\bf x})} \frac{\delta S_{\rm cl}}{\delta \phi(\bf x)}+
\frac{\delta S_{\rm eff}^{\rm 1-l}}{\delta \phi(\bf x)} =0
\end{equation}
where $S_{\rm cl}$ may be absent.
The one-loop effective action takes usually the form
\begin{equation}
S_{\rm eff}^{\rm 1-l}
 = \pm \frac{1}{2} \ln \frac{\det (-\Delta + m^2 + V(\phi(x)))}
{\det (-\Delta + m^2 )}
\end{equation}
where the sign $+$ or $-$ refers to bosonic and fermionic actions
respectively. The Klein-Gordon type operators will in general
be multichannel operators, mass term and potential being given by
matrices.
Two methods for computing the effective action numerically have been
proposed recently in [12,13] and in [14-19] 
and have been applied in various contexts [20-24].
Both methods are closely related. In both of them the usual problems of
renormalization and of removing zero and unstable modes are
easy to handle. While the second method , based on a theorem on
functional determinants \cite{CoIn},
is considerably faster the first of them
is more convenient in the present context. It can be adapted
easily for computing the functional derivative $\delta S_{\rm eff}^{
\rm 1-l}
/\delta \phi({\bf x})$. Renormalization remains easy to handle
and will be discussed.
We will not consider zero or unstable modes here. These are
particular to specific models and here we would like to keep
the presentation of the method general. For reasons of economy we
restrict ourselves furthermore to the case of one scalar external
field; this restriction can easily be removed. For
the fluctuating field we choose a Dirac field as the simplest
case of a coupled channel problem for the fluctuation operator.

\section{The fermion determinant and its functional derivative}
\setcounter{equation}{0}
We consider a fermion field in $4$ Euclidean dimensions which receives
its mass $m_F = g v$ from its
coupling to a Higgs field $\phi(x)$ which takes the
vacuum expectation value $v$.
The Dirac operator has the form
\begin{equation}
{\cal D} = \mbox{$\partial$\hspace{-0.55em}/}
- g \phi(x) = {\cal D}^{(0)} - g (\phi(x) -v ),
\end{equation}
where ${\cal D}^{(0)}= \mbox{$\partial$\hspace{-0.55em}/}
- m_F$ is the free Dirac operator
in the `broken symmetry phase'.
${\cal D}$ is the Euclidean Dirac operator, the gamma matrices
are hermitean and satisfy
$\{\gamma _\mu , \gamma _\nu \} = 2 \delta_{\mu\nu} $.
Since ${\cal D} $ is not a positive definite operator
we introduce the operator
\begin{equation}
{\cal M} = \gamma_5 {\cal D} \gamma_5 {\cal D} = -\partial^2 + m_F^2 +
{\cal V}(x),
\end{equation}
with the potential
\begin{equation}
\label{pot1}
{\cal V}(x) = g^2 ( \phi^2(x) - v^2)
+ g \mbox{$\partial$\hspace{-0.55em}/} \phi(x) \; .
\end{equation}
The Euclidean effective action induced by the fermion field
is then obtained as \footnote{We omit the superscript ${\rm 1-l}$
in the following.}
\begin{eqnarray}
S_{\rm eff}[\phi] &=& 
-\ln \det \left ( \frac{{\cal D}}{{\cal D}^{(0)}} \right ) = 
-\frac{1}{2} \ln \det \left( \frac{-\partial^2 + m_F^2 + {\cal V}(x)}
{-\partial^2 + m_F^2} \right)\nonumber\\
\label{lndet}
&=& -\frac{1}{2} \ln \det \left( 1 + 
(-\partial^2 + m_F^2)^{-1} {\cal V}(x) \right).
\end{eqnarray}

As a next step we will derive an explicit expression for the
functional derivative of $S_{\rm eff}[\phi]$ with respect to
$\phi(x)$. As a starting point we use the perturbative
expansion of the effective action with respect to the
potential ${\cal V}(x)$. We define the free Green function
${\cal G}_0(x-x')$ as the solution of the equation
\begin{equation}
(-\partial^2 + m_F^2) {\cal G}_0 (x-x') =  \delta^{(4)} (x-x')
\end{equation} 
where $\partial^2$ is the Euclidean Laplace operator in 4 dimensions.
Then the effective action can be expanded as
\begin{equation}
S_{\rm eff}[\phi] = \sum_{n=1}^{\infty}
\frac{(-1)^n}{2n} {\rm tr} \int \! \prod_{i=1}^{n} {\rm d}^4\!x_i \, 
{\cal G}_0(x_i
- x_{i-1}) \, {\cal V}(x_i) \qquad (\mbox{with } x_0:=x_n) \; .
\end{equation}
Obviously only ${\cal V}(x)$ depends on  $\phi(x)$, so the functional
derivative is obtained in a straightforward way as:
\begin{equation}
\frac{\delta S_{\rm eff}}{\delta \phi(x)} = \sum_{n=1}^{\infty}
\frac{(-1)^n}{2} {\rm tr}
\int d^4\!x_1 \frac{\delta {\cal V}(x_1)}{\delta
\phi(x)} \int \! \left ( \prod_{i=2}^{n} {\rm d}^4\!x_i \, 
{\cal G}_0(x_i-x_{i-1})
\, {\cal V}(x_i) \right) {\cal G}_0(x_1-x_n).
\end{equation}
If one looks for static solutions, the actual variation will be with
respect to fields depending only on the three dimensional vector
${\bf x}$. We will restrict our considerations to this case in the
following. For instanton computations (see e.g. \cite{BaaDai}) one
introduces an extra Euclidean
time variable, so that the present restriction
represents no loss of generality.
Obviously the time integrations can be carried out explicitly.
In order to do so we introduce the Green function
${\cal G}_0({\bf x}-{\bf x}',\nu)$ which is related to 
${\cal G}_0(x-x')$ via a Fourier
transform with respect to Euclidean time $\tau$. It satisfies
\begin{equation}
(\Delta - m_F^2-\nu^2) {\cal G}_0({\bf x}-{\bf x}',\nu) = 
- \delta^{(3)}({\bf x}-{\bf x}')
\end{equation}
and is given explicitly as
\begin{equation}
\label{g0fourwithkappa}
G_0({\bf x}-{\bf x}',\nu) = \int \! \frac{{\rm d}^3\!p_i}{(2\pi)^3} \,
\frac{\exp(i{\bf p}_i({\bf x}-{\bf x}'))}
{{\bf p}_i^2+\kappa^2}
\end{equation}
with $\kappa^2:=m_F^2+\nu^2$.
The time integration leads to delta functions for the differences
of the Euclidean energies $\nu_i$, leaving just one
integration with respect to $\nu$.
Explicitly we find
\begin{eqnarray}
\label{ablseff1}
\frac{\delta S_{\rm eff}}{\delta \phi({\bf x})}
&=& 2\pi \, \delta(0) \,
\left(-\frac{1}{2}\right){\rm tr} \int d^3\!x_1 \frac{\delta
{\cal V}(x_1)}{\delta \phi({\bf x})}\\
&\times& \int \limits_{-\infty}^{\infty} \! \frac{{\rm
d}\nu}{2\pi} \, \sum_{n=1}^{\infty} (-1)^{n+1} \left ( 
\prod_{i=2}^{n} {\rm d}^3\!x_i \, 
{\cal G}_0({\bf x}_i - {\bf x}_{i-1},\nu) \, {\cal V}({\bf x}_i)
\right) {\cal G}_0({\bf x}_1-{\bf x}_n,\nu).\nonumber
\end{eqnarray}
The factor $2 \pi \delta(0)$ is due to a delta function
$2 \pi \delta (\nu_n-\nu_0)$ whose argument vanishes due to the
cyclic identification of $\nu_n$ with $\nu_0$. It is to be identified,
as usual, with Euclidean time via
$2 \pi \delta(0) \to \tau$. Indeed, for time independent
external potentials the effective action is 
proportional to $\tau$ and in fact
we are considering $S_{\rm eff}/\tau$,
the zero point energy.
As $G_0({\bf x}-{\bf x}',\nu)$
is an even function of $\nu$ the integral
over this variable may be replaced as
\begin{equation}
\int \limits_{-\infty}^{\infty} {\rm d}\nu \longrightarrow 
2 \int \limits_{0}^{\infty} {\rm d}\nu.
\end{equation}
Furthermore the perturbative series
in the potential ${\cal V}({\bf x})$ can be
recollected into an exact Green function
\begin{equation} \label{greenexp}
{\cal G}({\bf x},{\bf x}',\nu) \sum_{n=1}^{\infty} (-1)^{n+1} \left ( \int
\prod_{i=2}^{n} {\rm d}^3\!x_i \,
{\cal G}_0({\bf x}_i - {\bf x}_{i-1},\nu) \, 
{\cal V}({\bf x}_i)
\right) {\cal G}_0({\bf x}'-{\bf x}_n,\nu)
\end{equation}
with ${\bf x}_1={\bf x}$.
${\cal G}({\bf x},{\bf x}',\nu)$ satisfies
\begin{equation}
\label{gdef}
\left( \Delta - \kappa^2 - {\cal V}({\bf x}) \right) 
{\cal G}({\bf x},{\bf x}',\nu)
= -\delta^{(3)}({\bf x}-{\bf x}').
\end{equation}
With these replacements the derivative of the effective action 
becomes
\begin{equation}
\label{ablseff}
\frac{1}{\tau}
\frac{\delta S_{\rm eff}}{\delta \phi({\bf x})} = - {\rm tr}
\int d^3\!x_1
\frac{\delta {\cal V}({\bf x}_1)}{\delta \phi({\bf x})} \int
\limits_{0}^{\infty} \!
\frac{{\rm d}\nu}{2\pi} \, {\cal G}({\bf x}_1,{\bf x}_1,\nu).
\end{equation}
It remains to evaluate the derivative of the potential 
${\cal V}({\bf x})$ with
respect to $\phi$. One obtains
\begin{eqnarray}
\frac{\delta {\cal V}
({\bf x}_1)}{\delta \phi({\bf x})} &=& \frac{\delta}{\delta
\phi({\bf x})} \left ( g^2 (\phi^2({\bf x}_1)-v^2) + g 
\mbox{\boldmath
$ \gamma \nabla \!$\unboldmath}_{{\bf x}_1} \phi({\bf x}_1)
\right)\nonumber\\
&=& \left ( 2g^2\phi({\bf x}_1) + 
g \mbox{\boldmath $\gamma \nabla \!
$\unboldmath}_{{\bf x}_1} \right)
\delta^{(3)}({\bf x}_1-{\bf x}).\nonumber
\end{eqnarray}
Inserting this expression into (\ref{ablseff}) yields the result
\begin{equation}\label{ablseff2}
\frac{1}{\tau} \frac{\delta S_{\rm eff}}{\delta \phi({\bf x})} -\frac{g^2 \phi({\bf x})}{\pi}
\int \limits_{0}^{\infty} {\rm d}\nu {\rm tr}
{\cal G}({\bf x},{\bf x},\nu) +
\frac{g}{2\pi} \int \limits_{0}^{\infty}
{\rm d}\nu {\rm tr} \mbox{\boldmath $ \gamma \nabla$\unboldmath}
 {\cal G}({\bf x},{\bf x},\nu).
\end{equation}
Here the gradient in the last term acts on the Green function
at equal arguments, i.e. one takes the gradient after taking the limit
of equal arguments.
The expression we have obtained 
is a formal one, yet. The Green functions are not defined
at equal arguments, they have to be regularized and a renormalization
has to be performed. This will be discussed below;  we prepare
this here, however, by introducing a convenient notation.
As we have seen  ${\cal G}({\bf x},{\bf x}',\nu)$ can be expanded
with respect to ${\cal V}({\bf x})$ via (\ref{greenexp}). The trace 
with respect to ${\bf x}$ and with respect to internal indices
is equivalent to a summation of one loop
Feynman graphs. In the following we denote
by a superscript $(k)$ a contribution  of order
$k$ in the potential ${\cal V}({\bf x})$ and by the  superscript
$\overline{(k)}$ the sum of all such contributions starting
with order $k$.
With this convention we can define a finite part of the functional
derivative via
\begin{equation}
\label{ablseffres}
 \displaystyle
\frac{1}{\tau} \frac{\delta S_{\rm eff}^{\overline{(3)}}}
{\delta \phi({\bf x})} -\frac{g^2 \phi({\bf x})}{\pi}
\int \limits_{0}^{\infty} {\rm d}\nu {\rm tr}
{\cal G}^{\overline{(2)}}({\bf x},{\bf x},\nu) +
\frac{g}{2\pi} \int \limits_{0}^{\infty} 
{\rm d}\nu {\rm tr} \mbox{\boldmath $\gamma \nabla$\unboldmath}
{\cal G}^{\overline{(2)}}({\bf x},{\bf x},\nu) .
\end{equation}
This quantity can be evaluated numerically; the first and second order
contributions are divergent and will be considered later.

In some applications the effective action is computed in a quantum
state in which some of the fluctuations are excited, e. g. in the form
of occupied valence quark states. Then the ``sea quark'' Green function
has to be modified by adding the ``valence quark'' contributions
\begin{equation}
\int\frac{d\nu}{2\pi}
{\cal G}^{\rm val}({\bf x},{\bf x},\nu) = \sum_{\alpha}
\frac{1}{2 E_\alpha} \psi_\alpha({\bf x})\psi_\alpha^\dagger({\bf x})
\end{equation}
where $\alpha$ runs over the occupied levels.
Since the operator $\mbox{\boldmath $\gamma \nabla$\unboldmath}
-g \phi({\bf x})$ occuring in (\ref{ablseff2}) is just the static
Dirac operator and
\begin{equation}
\left(-\mbox{\boldmath $\gamma \nabla$\unboldmath}+g \phi({\bf x})
\right)
 \psi_\alpha({\bf x}) E_\alpha \gamma_0 \psi_\alpha({\bf x})
\end{equation}
this means that we obtain an additional valence contribution
\begin{equation}
\frac{1}{\tau} \frac{\delta S_{\rm eff}^{\rm val}}
{\delta \phi({\bf x})}
=- g\sum_\alpha \bar\psi_\alpha({\bf x})\psi_\alpha
({\bf x})
\end{equation}
as to be expected.

\section{Computation of the Green function }
\setcounter{equation}{0}
The computation of the Green function ${\cal G}({\bf x},{\bf x}',\nu)$
has been considered previously in Ref. \cite{Baa1, Baa2}.
In order to proceed we have to introduce a further restriction
on the classical configuration $\phi({\bf x})$. We have to assume that
it is spherically symmetric. For most cases of interest
this is the case, as minimal or extremal solutions have usually
the highest possible symmetry.
Then we may expand the exact Green function into partial wave
Green functions. These can be obtained by solving the radial
Schr\"odinger equation numerically.
\subsection{Partial wave reduction}
For the partial wave reduction we have to use an appropriate 
spherical spinor basis. It can be found in
\cite{BjoDre}. One introduces the 2-spinors  
\begin{eqnarray}
\label{phiplus}
\varphi^{(+)}(\hat x) &=& \frac{1}{\sqrt{2l+1}}
\left ( 
	\begin{array}{l}
	\sqrt{l+1/2+m} \, Y_l^{m-1/2}(\hat x)\\
	\sqrt{l+1/2-m} \, Y_l^{m+1/2}(\hat x)
	\end{array}
\right) \quad \mbox{for } j=l+1/2,\\
\label{phiminus}
\varphi^{(-)}(\hat x) &=& \frac{1}{\sqrt{2l+1}}
\left ( 
	\begin{array}{l}
	\sqrt{l+1/2-m} \, Y_l^{m-1/2}(\hat x)\\
	-\sqrt{l+1/2+m} \, Y_l^{m+1/2}(\hat x)
	\end{array}
\right) \quad \mbox{for } j=l-1/2.
\end{eqnarray}
which are eigenspinors
of ${\bf J}^2,{\bf J}_z,{\bf L}^2$ and ${\bf S}^2$.
Further they satisfy
\begin{equation}
{\bf \sigma \hat x} \varphi^{(\pm) } = \varphi^{(\mp)}
\end{equation}
Using these spinors one can construct 4-spinors and reduce
the Dirac equation into two coupled radial equations
for the partial waves. 
In order to obtain a similar Ansatz for the Green function
we introduce the matrices
$\Omega_j^{(\pm)}(\hat x,\hat x^\prime)$ via
\begin{equation}
\label{omega}
\Omega_j^{(\pm)}(\hat x,\hat x^\prime) := \sum_{m=-j}^{j}
\varphi_{jm}^{(\pm)}
(\hat x) \varphi_{jm}^{(\pm) \dagger}(\hat x^\prime) ,
\end{equation}
in terms of which the Green function can be expanded as
\begin{equation}
{\cal G}^{(\pm)}({\bf x},{\bf x}^\prime,\nu)= \sum_{j}
\left(
\begin{array}{ll}
g_{j,11}^{(\pm)}
(r,r^\prime,\nu) \, \Omega_j^{(\pm)}(\hat x,\hat x^\prime)
& -g_{j,12}^{(\pm)}
(r,r^\prime,\nu) \, \Omega_j^{(\pm)}(\hat x,\hat x^\prime)
i\vec \sigma \hat x^\prime\\[2mm]
g_{j,21}^{(\pm)}(r,r^\prime,\nu) \, 
i\vec \sigma \hat x \Omega_j^{(\pm)}(\hat x,\hat x^\prime)
& g_{j,22}^{(\pm)}
(r,r^\prime,\nu) \, \Omega_j^{(\mp)}(\hat x,\hat x^\prime)
\end{array}
\right) \;.
\end{equation}
The partial wave Green functions $g_{j,nm}^{(\pm)}(r,r',\nu)$ satisfy
\begin{equation} 
\label{smallg}
\left ( \delta_{nk} (- D_n+\kappa^2) + v_{nk}(r) \right )
g_{km}^{(\pm)}(r,r^\prime,\nu) 
= \delta_{nm}
\frac{\delta(r-r^\prime)}{r^2} \; .
\end{equation}
Here $D_n$ is the radial operator
\begin{equation}
D_n = \frac{d^2}{dr^2}+\frac{2}{r}\frac{d}{dr} -
\frac{l_n(l_n+1)}{r^2}
\end{equation}
with $l_1=j-1/2$ and $l_2 = j+1/2$ for $g_j^{(+)}$ and
$l_1=j+1/2$ and $l_2=j-1/2$ for $g_j^{(-)}$.  
The potential has been reduced to a $(2\times 2)$ matrix
\begin{equation}
{\bf v(r)}= \left (
\begin{array}{ll}
g^2 \, (\phi^2-v^2) & -g \, \phi^\prime\\
-g \, \phi^\prime & g^2 \, (\phi^2-v^2)
\end{array}
\right ) \; .
\end{equation}
One sees easily that the $(+)$ and $(-)$ systems 
at equal $j$ are
equivalent upon identifying
\begin{equation}
g_{11}^{(-)}\to g_{22}^{(+)}, \quad g_{22}^{(-)}\to g_{11}^{(+)}, \quad
g_{12}^{(-)}\to g_{21}^{(+)}, \quad g_{21}^{(-)}\to g_{12}^{(+)}.
\end{equation}
\subsection{Partial wave Green functions}
The partial wave Green function can be expressed in standard way
by regular and singular solutions of the homogeneous
equation
\begin{equation}
\label{fdgl}
\sum_{k}\left [ \delta_{nk} (- D_n +\kappa^2 ) + 
v_{nk}(r) \right ] f_k(r) = 0 .
\end{equation}
This system has a fundamental set of two independent
two-component
 solutions regular as $r \to \infty$ which
will be denoted as $f^{\alpha +}_k$ ($\alpha=1,2$) and two
such solutions regular at $r=0$, denoted as  $f^{\alpha -}_k$.
In terms of these solutions
 the partial wave Green function can be expressed as
\cite{Baa2}
\footnote{We suppress the angular momentum index $j$, the superscript
$(\pm)$ and the argument $\nu$ in the following.}
\begin{equation} \label{basgreen1}
g_{km}(r,r^\prime)= \sum_{\alpha,\beta} \left \{
\Theta(r-r^\prime) f_k^{\alpha +}(r)
\omega_{\beta \alpha}^{-1} f_m^{\beta -}(r^\prime) + 
\Theta(r^\prime-r) f_k^{\alpha -}(r)
\omega_{\alpha \beta}^{-1} f_m^{\beta +}(r^\prime) \right \} .
\end{equation}
where $\omega_{\alpha \beta}$ is related to the Wronskians of the
two sets of solutions via
\begin{equation} \label{basgreen2}
\omega_{\alpha \beta}=r^2 \sum_n \left ( f_n^{\alpha+}\frac{d}{dr}
f_n^{\beta -}  - f_n^{\beta -}  \frac{d}{dr}
f_n^{\alpha +} \right ) = r^2 W(f^{\alpha+}(r),f^{\beta-}(r))\; .
\end{equation}
A short derivation of this formula is given in the Appendix.
It is convenient to choose the two sets of solutions in the form
\begin{equation} \label{fansatz}
f_n^{\alpha \pm} = \left (
\delta_n^{\alpha} + h_n^{\alpha\pm}(r)\right ) b_{l_n}^{\pm}(\kappa r)
\; .\end{equation}
Here $b_l^+(z) = k_l(z)$ and
$b_l^-(z) =i_l(z)$ are the modified spherical
Bessel functions regular for $ z \to \infty$ and as $ z \to 0$,
respectively. The functions $h_n^{\alpha\pm}$ are regular both at
$r= $ and as $r\to\infty$. They can be chosen to vanish
as $ r \to \infty$ as a boundary and normalization condition.
Then the Wronskian simplifies and we have
\begin{equation}
\omega_{\alpha\beta} = \delta_{\alpha\beta}/\kappa \;  .
\end{equation}
We will need the Green function only at $r=r'$. The limits
$r' \searrow r$ and $ r'\nearrow r $ are equal by construction,
they involve however different sets of functions. The numerical
differences are found to be unimportant.
We have chosen to take the symmetric limit
\begin{equation}
g_{mn}(r,r) = \sum_{\alpha} \frac{\kappa}{2} 
\left \{ f_m^{\alpha +}(r) f_n^{\alpha -}(r) + 
f_m^{\alpha -}(r)f_n^{\alpha +}(r) \right \} \; .
\end{equation}

\subsection{Removing the leading orders in the external potential}

As mentioned in section 2 we will have to remove the leading orders in
the external potential
${\cal V}({\bf x})$ in order to obtain the finite part 
$\delta S_{\rm eff}^{\overline{(3)}}/\delta\phi({\bf x})$; this implies 
obtaining the Green function in order $\overline{(2)}$. We note that
our ansatz (\ref{fansatz}) for the solutions $f_n^{\alpha \pm}$
represents a separation into the zeroth order part proportional
to $\delta_n^{\alpha}$ and the order $\overline{(1)}$ contribution
proportional to the functions $h_n^{\alpha \pm}$.
These functions satisfy the coupled differential equations
\begin{equation}
\label{hdgl}
 \displaystyle
\left( \frac{ d^2}{ dr^2} + 2\left( \frac{1}{r} +
\kappa \frac{b_{l_n}^{\prime \pm}}{b_{l_n}^\pm} 
\right) \frac{d}{dr} 
\right)h_n^{\alpha \pm}(r)
= \sum_{m}v_{nm}\left( \delta_m^\alpha + h_m^{\alpha
\pm} \right) \frac{b_{l_m}^\pm}{b_{l_n}^\pm} \; .
\end{equation}
which can be written in symbolic form as
\begin{equation}
{\cal D} h^\pm = V\,(1+h^\pm) \; .
\end{equation}
If one expands the functions $h$ with respect to powers of
$V$ as
\begin{equation}
h^\pm := h^{\pm \overline{(1)}} = h^{\pm (1)} +
h^{\pm (2)} + \dots ,
\end{equation}
one obtains the following sequence of equations
\begin{eqnarray}
\label{h1bdgl}
{\cal D} h^{\pm \overline{(1)}} &=& V\,\left(1+h^{\pm 
\overline{(1)}}\right),\\
\label{h1dgl}
{\cal D} h^{\pm (1)} &=& V,\\
\label{h2bdgl}
{\cal D} h^{\pm \overline{(2)}} &=& V\,h^{\pm \overline{(1)}}
\end{eqnarray}
which can be solved numerically. In terms of these functions the 
partial wave Green function at $r=r'$ can be expressed as
\begin{eqnarray}
g_{nm}(r,r) \\ &=&  
 \frac{\kappa}{2} \left
 \{ \left( \delta_m^n + h_m^{n+(1)} + h_n^{m-(1)} \right)
i_{l_n} k_{l_m} + \left( \delta_m^n + h_n^{m+(1)} + h_m^{n-(1)} \right)
i_{l_m} k_{l_n} \right.\nonumber\\
&+& \left( h_n^{m-\overline{(2)}} + h_m^{n+\overline{(2)}} +
\sum_{\alpha}h_m^{\alpha
+\overline{(1)}} h_n^{\alpha -\overline{(1)}} \right) i_{l_n}
k_{l_m}\nonumber\\
&+& \left. \left( h_n^{m+\overline{(2)}} + h_m^{n-\overline{(2)}} +
\sum_{\alpha}h_m^{\alpha -\overline{(1)}}
h_n^{\alpha +\overline{(1)}} \right) i_{l_m}
k_{l_n}\right\} \; .
\end{eqnarray}
The third and fourth term represent the contribution of order
$\overline{(2)}$
\begin{equation}
\begin{array}{rcl}
\displaystyle g_{mn}^{\overline{(2)}}(r,r) &=& \displaystyle
\frac{\kappa}{2} 
\left \{ \left( h_n^{m-\overline{(2)}} + h_m^{n+\overline{(2)}} +
\sum_{\alpha} h_m^{\alpha
+\overline{(1)}} h_n^{\alpha -\overline{(1)}} \right) i_{l_n}
k_{l_m} \right.\\
&+& \displaystyle
\left. \left( h_n^{m+\overline{(2)}} + h_m^{n-\overline{(2)}} +
\sum_{\alpha} h_m^{\alpha
-\overline{(1)}} h_n^{\alpha +\overline{(1)}} \right) i_{l_m}
k_{l_n}\right \}.
\end{array}
\label{g2b}
\end{equation}
\subsection{The derivative of the effective action}
Having at our disposal a numerical method to compute the
Green function it is straightforward, using
(\ref{ablseffres}), to derive an expression
which allows the numerical computation of the functional derivative
of the effective action. We first have to take some traces of the
Green function  and its derivatives.
Using the identities 
\begin{equation}
\label{omegaglg1}
\Omega_j^{(\pm)}(\hat x,\hat x) = \frac{j+1/2}{4\pi}
\left(
\begin{array}{rr}
1&0\\
0&1
\end{array}
\right) \;.
\end{equation}
and
\begin{eqnarray*}
&\sum \limits_{m=-l}^{l}& \vert Y_l^m \vert^2 = \frac{2l+1}{4\pi},\\
&\sum \limits_{m=-l}^{l}& m \vert Y_l^m \vert^2 = 0\\
 \qquad &\sum \limits_{m=-j}^{j}&
\sqrt{l+\frac{1}{2}-m} \; 
\sqrt{l+\frac{1}{2}+m} \; Y_l^{m+1/2} \, Y_l^{m-1/2 \ast}
= 0
\end{eqnarray*}
one finds
\begin{eqnarray}
\label{traceg}
{\rm tr} G^{\overline{(2)}}({\bf x},{\bf x},\nu) &=&
\frac{1}{4\pi} \sum_{j=1/2}^{\infty} (2j+1)
\left[ g_{j,11}^{\overline{(2)}}(r,r) + g_{j,22}^{\overline{(2)}}(r,r)
\right],\\
{\rm tr} \mbox{\boldmath $\gamma \nabla$ \unboldmath }
 G^{\overline{(2)}}({\bf x},{\bf x},\nu) &=&
-\frac{1}{4\pi}\sum_{j=1/2}^{\infty} 
2(2j+1) \left[ g_{j,12}^{\prime \overline{(2)}}(r,r) 
+ \frac{2}{r} \, g_{j,12}^{\overline{(2)}}(r,r) \right] .
\label{traceggg}
\end{eqnarray}
Supplying a factor of $2$ in order to take into account
both the $l=j+1/2$ and the $l=j-1/2$ modes
we obtain the final result
\begin{equation}
\label{endergebnis}
 \displaystyle
\begin{array}{rcl}
\displaystyle
\frac{1}{\tau} \frac{\delta S_{\rm eff}^{\overline{(3)}}}
{\delta \phi({\bf x})} &=& \displaystyle
-\frac{g^2 \phi(r)}{2\pi^2} \int \limits_{0}^{\infty} {\rm d}\nu 
\sum \limits_{j=1/2}^{\infty}
(2j+1) \left[ g_{j,11}^{\overline{(2)}}(r,r) 
+ g_{j,22}^{\overline{(2)}}(r,r)
\right]\\
& & \displaystyle -\frac{g}{2\pi^2} \int \limits_{0}^{\infty} 
{\rm d}\nu \sum \limits_{j=1/2}^{\infty} (2j+1) \left[ g_{j,12}^{\prime 
\overline{(2)}}(r,r) 
+ \frac{2}{r} \, g_{j,12}^{\overline{(2)}}(r,r) \right] .
\end{array}
\end{equation}
This expression is finite and can be computed numerically.
However, we have to include the renormalized divergent parts as well.
This will be done in the next section. Here we will discuss
some details of the numerical evaluation and we present an 
analytic approximation, based on the derivative expansion.
This analytic approximation will be compared to the numerical
results.
\subsection{Some details of the numerical evaluation}
The numerical evaluation of the effective action has been described
in extenso in previous publications \cite{Baa1,Baa2}. 
If one computes the effective action the first operation is to 
integrate the trace of the Green function
over the radial variable $r$. Thereby one obtains the
partial wave contributions at fixed $\nu$ which are then summed using
an asymptotic extrapolation. Finally one has to do the integration
over $\nu$. Here the information we want to draw from the Green 
function is a local one, we want to determine 
$\delta S_{\rm eff}/\delta \phi({\bf x})$ as 
a function of $r$.
This requires a different order of steps.
Let us write (\ref{endergebnis}) in the form
\begin{equation}
\frac{1}{\tau} \frac{\delta S_{\rm eff}^{\overline{(3)}}}
{\delta \phi({\bf x})} =- \frac{g}{2\pi^2} 
\int \limits_{0}^{\infty} {\rm d}\nu 
\sum \limits_{j=1/2}^{\infty} (2j+1) t_j(\nu,r)
\end{equation} 
where now
\begin{equation}
t_j(\nu,r)= g\phi(r)
\left[ g_{l,11}^{\overline{(2)}}(r,r) 
+ g_{j,22}^{\overline{(2)}}(r,r)
\right] + \left[ g_{j,12}^{\prime 
\overline{(2)}}(r,r) 
+ \frac{2}{r} \, g_{j,12}^{\overline{(2)}}(r,r) \right] .
\end{equation}
The fact that the method for computing the effective action
developed in \cite{Baa1,Baa2} is very fast is due to the fact that
after only two Runge-Kutte runs for
computing the functions $h_k^{\alpha \pm} (r)$ one knows
the partial wave Green function at all mesh points $r_n$. 
This advantage can be maintained
in the following way: the Green function is computed, at fixed
angular momentum $j$
for all $r_n$ in one sweep
and at each of the 
(roughly $2000$) mesh points $r_n$ the appropriate traces 
$t_j(\nu,r)$ are computed and stored as functions of 
$\nu$ and j. 

The computation was extended to $j - 1/2 \leq 25$ and
to $\nu \leq 2$. The partial wave sums 
$\sigma_n(\nu)  =\sum (2j+1) t_j(\nu,r_n)$ were computed using
an extrapolation of the form $a /j^4 + b/j^5$ to include the
terms with $j - 1/2 > 25$ and the integration over $\nu$ was 
extended beyond $\nu = 2$ by using an extrapolation of the
form $a/\nu^2 + b/ \nu^4$. The form of the extrapolations can
be derived rigorously, the convergence was checked for several values
of $r$.

The total time for one evaluation of $\delta S^{\overline{(3)}}
/\delta \varphi ({\bf x})$ at all values of $\vert {\bf x} \vert$
takes about  5 minutes on a workstation IBM Risc RS6000/340. 

\subsection{Analytic approximation}

In \cite{BaaSoSue} we presented an analytic approximation to
the finite part $S_{\rm eff}^{\overline{(3)}}$ of the action, using an
expansion with respect to gradients of the external
field $\phi$. It reads
\begin{eqnarray} \label{seff3ban} \nonumber
S_{\rm  eff}^{\overline{(3)}}[\phi] &\approx& \displaystyle
-\frac{g^4}{16\pi^2}
\int d^4\!x \left [
\phi^4 \ln \frac{\phi^2}{v^2} - \frac{3}{2}(\phi^2-v^2)^2 - 
v^2(\phi^2-v^2) \right ]\\
& &  \displaystyle
-\frac{g^2}{16\pi^2} \int d^4\!x (\partial_\mu \phi)^2 \left [ \ln
\frac{\phi^2}{v^2} - \frac{2}{3v^2}(\phi^2-v^2) \right] \; .
\end{eqnarray}
This approximation was found to be excellent at
sufficiently high fermion masses. It is interesting to compare
exact results and this analytic approximation also on the level
of the functional derivative. Taking the fields to be time independent
and dependent only on $r=|{\bf x}|$ we obtain
\begin{eqnarray} 
\frac{1}{\tau} \frac{\delta S_{\rm eff}^{\overline{(3)}}}{\delta 
\phi(\bf x)} &=& - \frac{g^4 \phi}{4\pi^2} \left[
\phi^2 \ln \frac{\phi^2}{v^2} - \phi^2 + v^2 \right]
- \frac{g^2}{16\pi^2} \left[ -(\phi^\prime)^2
\left( \frac{2}{\phi}
- \frac{4}{3v^2} \phi \right) \right.\nonumber\\
& & \left. {} - 2 \left( \phi^{\prime \prime}
+ \frac{2}{r} \phi^\prime \right)
\left( \ln\frac{\phi^2}{v^2} - \frac{2}{3v^2}
\left(\phi^2 -v^2 \right) \right) \right].
\label{ablseffanalyt}
\end{eqnarray}
This analytic approximation is compared to the numerical results
in Figs. 1 to 4. For the numerical evaluation we relate the physical
quantities to dimensionles ones by
$x=\hat x/v$ and $\phi=v\hat\phi$.
 Here and in the following we use for $\hat\phi$
the profile function
\begin{equation} \label{profile}
f(r) =(1+\exp(-\sqrt{r^2+1}+\sqrt{R^2+1}))^{-1}
\end{equation} 
with $R=2$ as in \cite{BaaSoSue}. It has the typical shape of a 
medium-sized bubble. The approximation is found to 
reproduce the numerical results quite well;
as to be expected the approximation improves with increasing
fermion mass. 

To the finite part of order $\overline{(3)}$ we have to add the
finite parts of the first and second order contributions. They
are known analytically and their numerical evaluation, involving
only a numerical Fourier transform and simple integrations, can
be performed with high precision. For this reason we have not 
included them here into the comparison between numerical and
approximate analytical results.

\section{Renormalization of the effective action}
\setcounter{equation}{0}
Having discussed the computation of the finite contributions
of third and higher order in the external potential
${\cal V}({\bf x})$ we have to come back
now to the computation of the first and second order contributions to
the effective action.
These contributions are divergent. As a first step
they have to be regularized, e. g. by dimensional or
Pauli-Villars regularization, introducing
thereby an ultraviolett cutoff $\Lambda$. If there
is no genuine action for the field $\phi$,
as e.g. in Nambu-Jona-Lasinio
type models where $\phi$ parametrizes fermion condensates, 
the theory will depend on this cutoff - the latter will be a
physically significant quantity. If the $\phi$ field has a genuine
action of the $\phi^4$ type then the divergences can be absorbed into
counterterms and the cutoff dependence disappears via renormalization
of the physical masses and couplings. We will consider the latter case
here; to be specific we choose a Higgs type potential for the
field $\phi$.

The bare Lagrangian of a Higgs-Yukawa theory reads
\begin{equation}
{\cal L} = \frac{1}{2} \partial_\mu \phi \partial^\mu \phi -
U(\phi) 
+ \bar \psi  \left( i \mbox{$\partial$\hspace{-0.55em}/}
 - g \phi \right) \psi \;.
\end{equation}
with
\begin{equation} \label{Higpot}
U(\phi) = \frac{\lambda}{4} \left( \phi^2-v^2 \right)^2 \;.
\end{equation}
If the Higgs field is assumed to take its vacuum expectation value
$v$ at large $r$ it is convenient to introduce
the field shift
\begin{equation}
\phi(x) = v + \varphi(x)
\end{equation}
and to express the Lagrangian in terms of $\varphi$ as
\begin{equation}
{\cal L} \frac{1}{2} \partial_\mu \varphi \partial^\mu \varphi
-\frac{1}{2} m_\varphi^2 \varphi^2 -\lambda v \varphi^3
-\frac{\lambda}{4} \varphi^4
+\bar\psi \left(i \mbox{$\partial$\hspace{-0.55em}/}
-(m_F +g\varphi)\right)\psi.
\end{equation}
Here we have introduced the boson mass $m_\varphi=\sqrt{2\lambda} v$
and the fermion mass $m_F= gv$.
\\ \\
The counterterm
Lagrangian will have to include a  fourth order
polynomial in the field
$\varphi$ and also a wave function renormalization
counterterm. We do not have to include here corresponding
counterterms for the fermion part of the action. Such terms would be
made necessary only if boson loops were included. The counterterm
Lagrangian reads then
\begin{equation}
\label{counterterme}
{\cal L}_{\rm c.t.}(\varphi) = 
-A \varphi - \frac{1}{2} B \varphi^2 - \frac{1}{6}
C \varphi^3 - \frac{1}{24} D \varphi^4
+ \frac{1}{2} \, \delta \! Z \, \partial_\mu \varphi
\partial^\mu \varphi
\end{equation}
For the Yukawa - Higgs Lagrangian we note that the bare Lagrangian 
depends only on the sqare of the unshifted field $\phi^2$ and its
derivative. This holds also true for the divergent terms of the
one-loop effective action generated by the fermion field, due to the
structure (\ref{Higpot}) of the external potential ${{\cal V}}(x)$.
Therefore, only two of the four constants $A,B,C$ and $D$ are 
independent. This implies the relations
\begin{eqnarray} \label{constraints}
A - Bv + \frac{1}{3}D v^3 = 0 \nonumber \\
C=vD.
\end{eqnarray}
The choice of
the independent quantities depends on the renormalization
conditions which will be given below.
\\ \\
The first and
second order contributions to the one-loop effective action
are given - first as formal definitions and then in dimensionally
regularized form - by
\begin{eqnarray} \label{divcon}
S_{\rm eff}^{(1)} &=& -\frac{1}{2} {\rm tr} \int d^4\!x {\cal G}_0(0)
{\cal V}(x) \nonumber\\
&=& \frac{2m_F^2}{(4\pi)^2} \left\{ 
\frac{1}{\varepsilon} +
\psi(1) + 1 +\ln \frac{4\pi\mu^2}{m_F^2} \right\}
\int d^4\!x (g^2 \varphi^2 + 2m_Fg\varphi),\nonumber\\
S_{\rm eff}^{(2)} &=& \frac{1}{4} {\rm tr} \int \! {\rm d}^4\!x
\, {\rm d}^4\!y \, {\cal G}_0(x-y) {\cal V}(y) \; {\cal G}_0(y-x)
{\cal V}(x)\nonumber\\
\nonumber \\ \label{Seff2}
&=& \frac{1}{64\pi^2} \int\frac{d^4q}{(2\pi)^4}
{\rm tr} \vert \widetilde {\cal V}(q) \vert^2 \left \{
 \frac{1}{\varepsilon} + \psi(1) + 
\ln \frac{4\pi\mu^2}{m_F^2} \right.\\ \nonumber
& & \left. {}+ 2 - \frac{1}{\rho(q)} \ln
\frac{1+\rho(q)}{1-\rho(q)} \right \}.
\end{eqnarray}
Here $\rho(q)$ is defined as $\rho(q):=q/\sqrt{q^2 + 4m_F^2}$.
We choose the renormalization conditions in analogy to 
Coleman und Weinberg \cite{CoWe}.
In order to do so we introduce at first
the one-loop effective potential. It is
given essentially by the one-loop action induced by constant fields
$\varphi(x)=\bar \varphi$. Then the external potential takes the form
\begin{equation}
\widetilde {\cal V}(q)g^2\left \{\begin{array}{cc} \bar \phi^2 -v^2 & 0
\\ 0 &\bar \phi^2 -v^2\end{array} \right \} \delta^4(q)
=\hat V (\bar\varphi) \left \{\begin{array}{cc} 1 & 0
\\ 0 &1\end{array} \right \} \delta^4(q)
\end{equation}
with
\begin{equation} \label{vhatdef}
\hat V(\varphi)= g^2 (\varphi^2 + 2 v\varphi) 
\end{equation}
and the effective action can be written as
\begin{equation}
S_{\rm eff}
[\varphi(x)=\bar\varphi]=  V_{\rm  eff}^{\rm 1-l} \Omega^{(4)}
\end{equation}
where $\Omega^{(4)} $ is the volume of space-time. 
The renormalized effective potential including the bare potential and
the counterterms is given by
\begin{equation}
V_{\rm eff,ren}:U(\phi=v+\bar \varphi) + V_{\rm eff}^{\rm 1-l} + \left
\{ A \bar\varphi + \frac{1}{2} B \bar\varphi^2 + \frac{1}{6}
C \bar\varphi^3 + \frac{1}{24} D \bar\varphi^4 \right\}
\; . \end{equation}
Here $V_{\rm eff}^{\rm 1-l}$ includes the one loop contributions to
all orders,
i. e. the regularized divergent parts of first and second order
and the $\overline{(3)}$ contribution of all higher orders.
The first and second order terms can be found easily 
from (\ref{divcon}); they read
\begin{equation}
\label{veff1l1reg}
V_{\rm eff, reg}^{\rm 1-l (1)} = 2 
\hat V(\bar\varphi)
\frac{m_F^2}{(4\pi)^2} \left( \frac{1}{\varepsilon}
 + \psi(1) + 1 + \ln \frac{4\pi\mu^2}{m_F^2}\right).
\end{equation}
and
\begin{equation} \label{veff1l2reg}
V_{\rm eff,reg}^{\rm 1-l (2)}
= \frac{\hat V^2(\bar\varphi)}{(4\pi)^2}
 \left( \frac{1}{\varepsilon} + \psi(1) + \ln
\frac{4\pi\mu^2}{m_F^2} \right).
\end{equation}
The $\overline{(3)}$ contribution has been given in
\cite{BaaSoSue}, it reads
\begin{eqnarray}
V_{\rm eff}^{\rm 1-l \overline{(3)}} &=& -2
\int\frac{d^4p}{(2\pi)^4}
\sum_{n=3}^{\infty} \frac{(-1)^{n+1}}{n} \left( \frac{\hat
V(\bar\varphi)}{p^2+m_F^2} \right)^n\nonumber\\
&=& - \frac{2 m_F^4}{(4\pi)^2} \left\{ \ln\left( 1+ \frac{\hat
V(\bar\varphi)}{m_F^2} \right) \left[ \frac{1}{2} + \frac{\hat
V(\bar\varphi)}{m_F^2} 
+ \frac{1}{2}\frac{\hat V^2(\bar\varphi)}{m_F^4} \right]
\right.\\
& & \left. {}-
\frac{\hat V(\bar\varphi)}{2m_F^2} - \frac{3}{4}\frac{\hat
V^2(\bar\varphi)}{m_F^4} \right\}\nonumber \;.
\end{eqnarray}
We require the effective potential 
to retain the following properties of the tree-level potential
$U(\phi=v+\varphi)$ :
\begin{eqnarray}
\left. \frac{{\rm d}^2 V_{\rm eff,ren}(\bar\varphi)}{{\rm d}
\bar\varphi^2} \right
\vert_{\bar\varphi=0} &=& m_\varphi^2 \qquad \mbox{(physical mass),}\\
\left. \frac{{\rm d}V_{\rm eff,ren}(\bar\varphi)}{{\rm d}\bar\varphi} 
\right \vert_{\varphi=0} &=& 0 \qquad (\langle \bar\varphi \rangle =0).
\end{eqnarray}
The first of these conditions fixes the constant $B$ of the counterterm
action, the second one the constant $A$. The remaining constants
are fixed by the constraints (\ref{constraints}).
One finds after some algebra
\begin{eqnarray*}
A &=& -\frac{gm_F^3}{4\pi^2} \left( \frac{1}{\varepsilon} + \psi(1) + 1
+ \ln \frac{4\pi\mu^2}{m_F^2} \right),\\
B &=& -\frac{3g^2m_F^2}{4\pi^2}
\left( \frac{1}{\varepsilon} + \psi(1) +
\frac{1}{3} + \ln \frac{4\pi\mu^2}{m_F^2}  \right),\\
C &=& -\frac{3g^3m_F}{2\pi^2} \left( \frac{1}{\varepsilon} + \psi(1) + 
\ln \frac{4\pi\mu^2}{m_F^2} \right),\\
D &=& -\frac{3g^4}{2\pi^2} \left( \frac{1}{\varepsilon} + \psi(1) + 
\ln \frac{4\pi\mu^2}{m_F^2} \right).
\end{eqnarray*}
With these counterterms the total effective potential
becomes
\begin{eqnarray} \label{vefftot}
V_{\rm eff,ren}(\varphi) &=& 
\frac{gm_F^3}{2(2\pi)^2} \varphi + \frac{1}{2} 
m_\varphi^2 \left( 1 + \frac{7g^2m_F^2}{2(2\pi)^2 m_\varphi^2} \right) 
\varphi^2\\ \nonumber
& & {}+ \lambda v
\left( 1+ \frac{3}{2(2\pi)^2} \frac{g^3m_F}{\lambda v} 
\right) \varphi^3 + \frac{1}{4} \lambda \left( 1+ \frac{3}{2(2\pi)^2} 
\frac{g^4}{\lambda} \right) \varphi^4\\ \nonumber
& & {}-\frac{2}
{(4\pi)^2} (g\varphi + m_F)^4 \ln \frac{g\varphi + m_F}{m_F}.
\end{eqnarray}
and is independent of $\varepsilon$ and $\mu$.
In order to determine the wave function renormalization we have
to consider $S_{\rm eff}^{(2)}$ again. Inserting the explicit
expression
for ${\rm tr}\vert(\tilde{\cal V})(q))\vert^2$ into (\ref{Seff2})
we find 
\begin{eqnarray}
\nonumber
S_{\rm eff}^{(2)} &=& 
\frac{1}{16\pi^2} \int\frac{d^4q}{(2\pi)^4}
\left( \vert\tilde {\hat V}(\varphi)(q)\vert^2 + 
g^2 q^2 \vert\tilde \varphi(q)\vert^2 \right)
\left \{ \frac{1}{\varepsilon} + \psi(1) + 
\ln \frac{4\pi\mu^2}{m_F^2} \right.\\ 
& & \left. {}+ 2 - \frac{1}{\rho(q)} \ln
\frac{1+\rho(q)}{1-\rho(q)} \right \}.
\end{eqnarray}
where $\hat V (\varphi)$ has been defined in (\ref{vhatdef}).
In the integrand the contribution
\begin{equation}
\vert\tilde{\hat V}
(\varphi)(q)\vert^2\left(\frac{1}{\varepsilon} + \psi(1) + 
\ln \frac{4\pi\mu^2}{m_F^2} \right)
\end{equation}
has been absorbed already into the renormalized effective potential.
(see (\ref{veff1l2reg}) and (\ref{vefftot})).
The contribution
\begin{eqnarray} \label{deltaZ1}
&&\frac{1}{16\pi^2} \int\frac{d^4q}{(2\pi)^4}
g^2 q^2 \vert\tilde \varphi(q)\vert^2 
\left(\frac{1}{\varepsilon} + \psi(1) + 
\ln \frac{4\pi\mu^2}{m_F^2}\right)
\\ \nonumber&&\frac{g^2}{16\pi^2}
\left(\frac{1}{\varepsilon} + \psi(1) + 
\ln \frac{4\pi\mu^2}{m_F^2}\right)\int d^4x (\partial_\mu \phi(x))^2
\end{eqnarray} 
will be absorbed into the wave function renormalization $\delta Z$
(see below). This leaves a finite part
\begin{eqnarray} \label{seff2fin}
S_{\rm eff,finite}^{(2)} &=& 
\frac{1}{16\pi^2} \int\frac{d^4q}{(2\pi)^4}
\left( \vert\tilde{\hat V}(\varphi)(q)\vert^2 + 
g^2 q^2 \vert\tilde \phi(q)\vert^2 \right)
\\ \nonumber &&\left \{2 - \frac{1}{\rho(q)} \ln
\frac{1+\rho(q)}{1-\rho(q)} \right \}.
\end{eqnarray}
The expression in curly brackets can be expanded as
\begin{equation}
2 - \frac{1}{\rho(q)} \ln
\frac{1+\rho(q)}{1-\rho(q)} \simeq -\frac{q^2}{6m_F^2} + O(q^4)
\; .\end{equation}
Therefore the term $g^2 q^2 \vert\tilde \phi(q)\vert^2$ 
in the first parenthesis
will be multiplied
at least with $q^2$ and yield terms with at least
four derivatives of the fields. On the other hand
the term $\vert(\tilde{\hat V}(\phi)(q)\vert^2$ will yield another 
second derivative term
when multiplied with $q^2/6m_F^2$. We denote this contribution as
$S_{\rm eff, analytic}^{(2)}$. It can be rewritten as
\begin{eqnarray} 
S_{\rm eff, analytic}^{(2)}
&=& -\frac{1}{16 \pi^2}\\int\frac{d^4q}{(2\pi)^4}
\vert\tilde{\hat V}(\varphi)(q)\vert^2\frac{q^2}{6m_F^2} \\ \nonumber
&=& -\frac{1}{24 \pi^2} \int d^4x\left ( \frac{g^2}{m_F^2}\varphi^2
+g^2+2\frac{g^3}{m_F}\varphi\right) (\partial_\mu \varphi)^2
\; .\end{eqnarray}
Since we renormalize at $\phi=v$, i. e. $\varphi=0$ we find here
another contribution to the wave function renormalization :
\begin{equation} \label{deltaZ2}
-\frac{g^2}{24 \pi^2} \int d^4x (\partial_\mu \varphi)^2 \; ,
\end{equation}
to be absorbed into $\delta Z$. Other second derivative terms appear in
$S_{\rm eff}^{\overline{(3)}}$ given in (\ref{seff3ban}), their
coefficient vanishes at $\phi=v$.
We find therefore, collecting the contributions (\ref{deltaZ1})
and (\ref{deltaZ2})
\begin{equation}
 \delta Z = 
-\frac{2g^2}{(4\pi)^2}\left(\frac{1}{\varepsilon} + \psi(1) + 
\ln \frac{4\pi\mu^2}{m_F^2}\right)+ \frac{g^2}{12\pi^2}
\; 
.\end{equation}
The effective action
\begin{equation}
\Gamma=S_{\rm cl}+S_{\rm eff}^{\rm 1-l}
\end{equation}
becomes therefore
\begin{equation}
\Gamma
= \int d^4 x \left(\frac{1}{2} (1+\frac{g^2}{12 \pi^2})(\partial_\mu
\phi)^2
+\frac{\lambda}{4} (\phi^2-v^2)^2 \right)
+S_{\rm eff,finite}^{(2)} + S_{\rm eff}^{\overline {(3)}}
\; .\end{equation}
While taking the functional derivative of the first term is trivial
and the numerical
computation of the term $S_{\rm eff}^{\overline {(3)}}$
has been discribed in section 2, we have to discuss, as a final step,
the computation of the functional derivative of 
$S_{\rm eff,finite}^{(2)}$. Restricting ourselves again
to time-independent configurations we find,
using the definition (\ref{seff2fin}),
\begin{eqnarray} \label{dseff2fin}
\nonumber &&\frac{1}{\tau}
\frac{\delta S_{\rm eff,finite}^{(2)}}{\delta\phi({\bf x})}
= \frac{g^2}{4\pi^2}
\left(2g^2 \phi({\bf x}) (\phi^2({\bf x})-v^2) -
\Delta \phi({\bf x})\right )
-\frac{g^2}{8\pi^2}\int\frac{d^3q}{(2\pi)^3}\exp(i{\bf q}{\bf x})
\\  &&
\left ( 2g^2\phi({\bf x})(\phi^2-v^2)^\sim (q)+
q^2(\phi-v)^\sim({\bf q}) \right )
\frac{1}{\rho(q)}\ln \frac{1+\rho(q)}{1-\rho(q)}
\; .\end{eqnarray}
Here $(..)^\sim$ denotes the Fourier transform of the expression in
parentheses. Taking into account of the fact that the fields depend
only on
$r=\vert{\bf x}\vert$ and therefore their Fourier transforms only
on $q=\vert{\bf q}\vert$, this expression can be further simplified.
Its computation involves simple one-dimensional
Fourier-Bessel transforms and integrations.

We compare the numerical results for the renormalized second order
contribution
$(1/\tau)\delta S_{\rm eff,finite}^{(2)}/\delta\phi({\bf x})
- g^2 \Delta \phi /12 \pi^2$
with those for $(1/\tau)\delta S^{\overline{(3)}}
/\delta \phi$ in Figure 5. The renormalized second order
contribution is found to be considerably smaller than the contributions
of order $\overline{(3)}$.    

\section{Conclusions}
We have presented here a method of computing the functional
derivative of an exact effective action $\Gamma$ for spherically
symmetric background field configurations. The vanishing of this
functional derivative determines field configurations that
minimize the action or, for time independent configurations,
the energy of a quantum system in one-loop order.
We have considered the exact computation of the finite part as
well as the implementation of renormalization, whenever it is
necessary. We have considered here a fermionic loop contribution,
the method is of course more general.

For the special case considered here we have also presented an 
analytic estimate 
\begin{eqnarray} 
\frac{1}{\tau} \frac{\delta S_{\rm eff}^{\overline{(3)}}}{\delta 
\phi({\bf x})} &=& - \frac{m_F^4 \phi}{4\pi^2v} \left[
\hat\phi^2\left( \ln \hat\phi^2 -1 \right) + 1 \right]
- \frac{m_F^2}{16\pi^2v}
\left[ -(\hat\phi^\prime)^2\left( \frac{2}{\hat\phi}
- \frac{4}{3} \hat\phi \right) \right.\nonumber\\
& & \left. {} - 2 \left(\hat \phi^{\prime \prime}
+ \frac{2}{r}
\hat\phi^\prime \right) \left( \ln\hat\phi^2 - \frac{2}{3}
\left( \hat \phi^2 -1 \right) \right) \right]
\end{eqnarray}
for the finite part of order $\overline{(3)}$ and compared it to
the exact numerical results. The approximation is found to be 
good at large $m_F$, as to expected. This local expression, as well as
some local terms occuring in 
$\delta S_{\rm eff, finite}^{(2)}/\delta \phi({\bf x})$
(see (\ref{dseff2fin})) can be 
introduced explicitly into the classical equation of motion 
(\ref{eqmot}) thus obtaining an improved differential
equation for the classical solution. Its solution should represent 
a good starting
point for an iterative procedure which takes into account
the remaining parts that have to be computed numerically. 
\newpage
\section*{Appendix}
\setcounter{equation}{0}
\renewcommand{\theequation}{A.\arabic{equation}}
Equations (\ref{basgreen1}) and (\ref{basgreen2}) are relations
 one expects to find
in some
textbook, therefore in \cite{Baa2} they were given without proof.
Since we have not found such a textbook yet, we give here a 
heuristic derivation.
The Green function we want to determine
satisfies
\begin{equation}
\left (\delta_{nk} (-\frac{d^2}{dr^2}
-\frac{2}{r}\frac{d}{dr} +\frac{l_n(l_n+1)}{r^2}
+\kappa^2) + v_{nk}(r) \right ) g_{km}(r,r')= \frac{1}{r^2}\delta(r-r')
\; .
\end{equation}
{}From the the fact that it is a solution of the homogeneous
differential equation for $r \neq r'$, both as a function
of $r$ and $r'$, and imposing regularity
at $r =0$ and $\infty$ we conclude that it must have the form
\begin{equation}
g_{mn}(r,r^\prime)= \sum_{\alpha,\beta} \left \{
\Theta(r-r^\prime) f_m^{\alpha +}(r)
c_{\alpha \beta} f_n^{\beta -}(r^\prime) + 
\Theta(r^\prime-r) f_m^{\alpha -}(r)
d_{\alpha \beta} f_n^{\beta +}(r^\prime) \right \}
\end{equation}
in terms of the functions $f^{\alpha\pm}_n(r)$ defined in section
3.2 . The Green function has to be continuous at $r=r'$. Its first
derivative has to have a discontinuity so that the second derivative
produces the appropriate $\delta$ function. 
These two conditions lead to the relations
\begin{eqnarray}
\sum_{\alpha,\beta}\left( c_{\alpha\beta}
f_m^{\alpha+}(r)f_n^{\beta-}(r)
-d_{\alpha\beta}f_m^{\alpha-}(r)f_n^{\beta+}(r)\right)&=&0 \\ 
r^2\sum_{\alpha,\beta} \left(c_{\alpha\beta}
f_m^{\prime\alpha+}(r)f_n^{\beta-}(r)
-d_{\alpha\beta} f_m^{\prime\alpha-}(r)
f_n^{\beta+}(r)\right) &=& - \delta_{mn}
\; .\end{eqnarray}
We contract the
first of these equations with $r^2f_m^{\prime\gamma+}(r)$
and the second
one with $f_m^{\gamma+}(r)$ , then we subtract the resulting
equations from each other. 
We obtain
\begin{equation}
\sum_{\alpha,\beta} \left(c_{\alpha\beta}
r^2 W(f^{\gamma+},f^{\alpha+})f_n^{\beta-}
-d_{\alpha\beta} r^2 
W(f^{\gamma+},f^{\alpha-})f_n^{\beta+}(r)\right) = - f_n^{\gamma+}(r)
\end{equation}
where we have defined the Wronskians as
\begin{equation}
W(f^{\alpha},g^{\beta})= \sum_m \left(f_m^{\alpha}(r)
g_m^{\prime \beta}(r)-f_m^{\prime\alpha(r)}g_m^{\beta}(r)\right)
\; .\end{equation}
If the potential $v_{nk}(r)$ is symmetric
the differential equation satisfied by the
functions $f^{\alpha\pm}(r)$ implies that 
$r^2 W(f^{\alpha},g^{\beta})$ is independent of $r$.
Since the functions $f^{\alpha+}_n(r)$ vanish as $r \to \infty$ the 
Wronskians $W(f^{\gamma+},f^{\alpha+})$ vanish.
Using the fact that the solutions $f^{\gamma+}(r)$ 
have been chosen to form a linearly independent set,
we conclude that
\begin{equation}
\sum_{\alpha}d_{\alpha\beta}
 r^2 W(f^{\gamma+},f^{\alpha-})
=\sum_{\alpha}d_{\alpha\beta}\omega^{\gamma\alpha} =\delta_\beta^\gamma
\end{equation}
or
\begin{equation}
d_{\alpha\beta} = \omega^{-1}_{\alpha\beta}
\;. \end{equation}
In order to obtain the coefficients $c_{\alpha\beta}$ we contract
(A.3) with $r^2f_m^{\prime\gamma-}(r)$ and (A.4) with 
$f_m^{\gamma-}(r)$. Using the fact that 
$r^2 W(f^{\gamma-},f^{\alpha-})$ vanishes because it does so 
as $r \to 0$, we find
\begin{equation}
c_{\alpha\beta} = \omega_{\beta\alpha}^{-1} \; .
\end{equation}
This closes
the demonstration of (\ref{basgreen1}) and (\ref{basgreen2}).

\newpage
\section*{Figure Captions}
{\bf Fig.~1} The
functional derivative of the finite part of the effective
action $(1/\tau)\delta S_{\rm eff}^{\overline{(3)}}/
\delta \phi({\bf x})$ in units $v^3$,
 calculated for a spherically symmetric time independent
profile $\phi({\bf x})=f(r)$ (see (\ref{profile})) with $R=2$. Here we
choose the fermion mass in units of $v$ to be $\hat m_F\equiv g=1.5$.
The solid line is the exact numerical
result and the dashed line is the analytic estimate given in Eq.
(\ref{ablseffanalyt}).\bigskip

\noindent
{\bf Fig.~2} The same as in Figure 1 for $g=1.7$.\bigskip

\noindent
{\bf Fig.~3} The same as in Figure 1 for $g=2.0$.\bigskip

\noindent
{\bf Fig.~4} The same as in Figure 1 for $g=2.2$.\bigskip

\noindent
{\bf Fig.~5}
The various contributions to the full one-loop correction of
the functional derivative of the effective action in units of $v^3$
for $g=2.0$:
The dotted line
represents the renormalized second order
contribution $(1/\tau)\delta S_{\rm eff,finite}^{(2)}/
\delta\phi({\bf x}) - g^2 \Delta \phi /12 \pi^2$; the dashed line
is the exact numerical result of
$(1/\tau)\delta S_{\rm eff}^{\overline{(3)}}
/\delta \phi$ which is also presented in Figure~3. The solid line
is the sum of both contributions.

\newpage
\begin{figure}
\centerline{
\mbox{\epsfxsize=16cm
\epsfbox{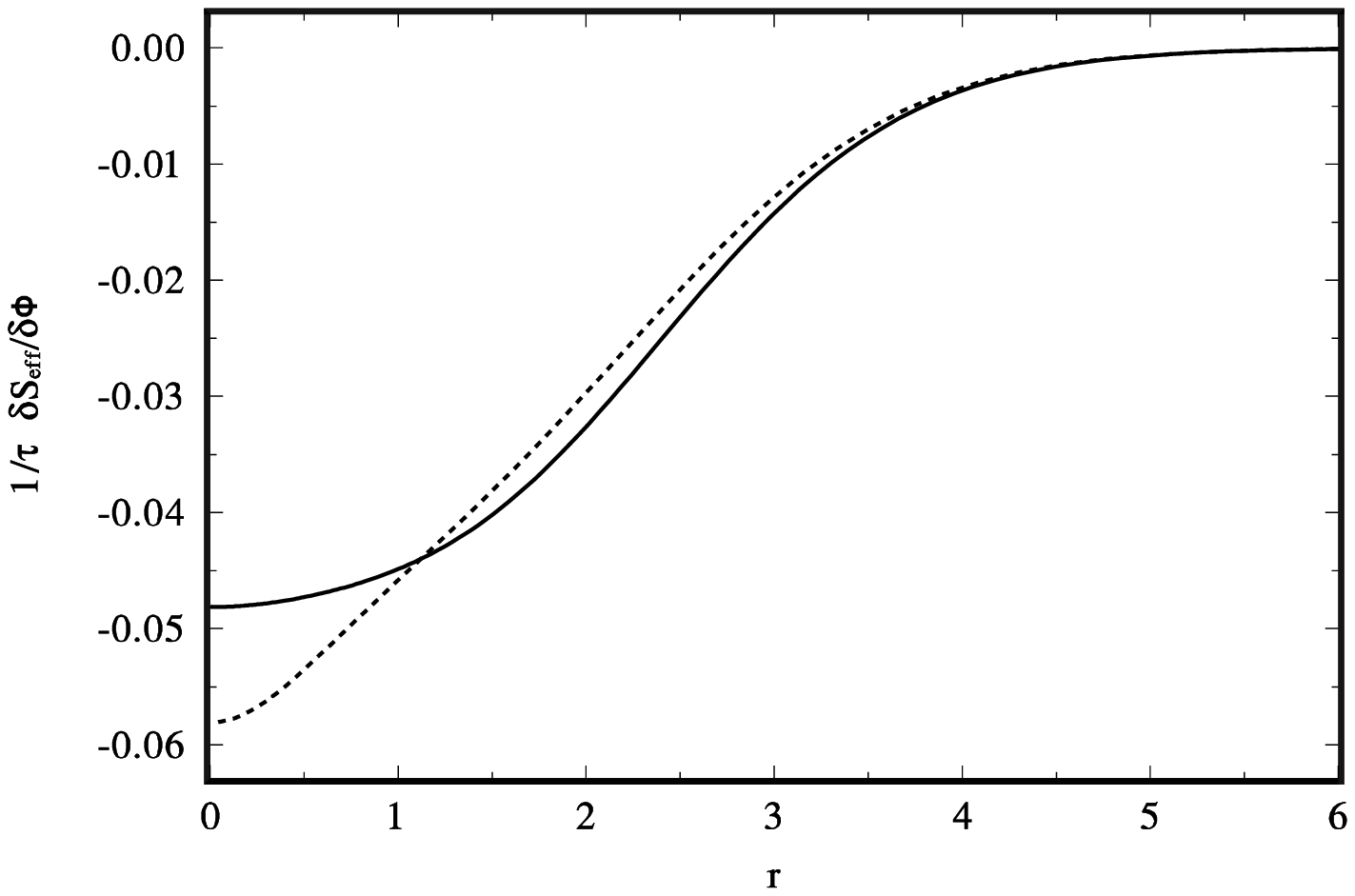}}
}

\vspace{3mm}

\centerline{\Large Figure 1}
\end{figure}
\clearpage
\begin{figure}
\centerline{
\mbox{\epsfxsize=16cm
\epsfbox{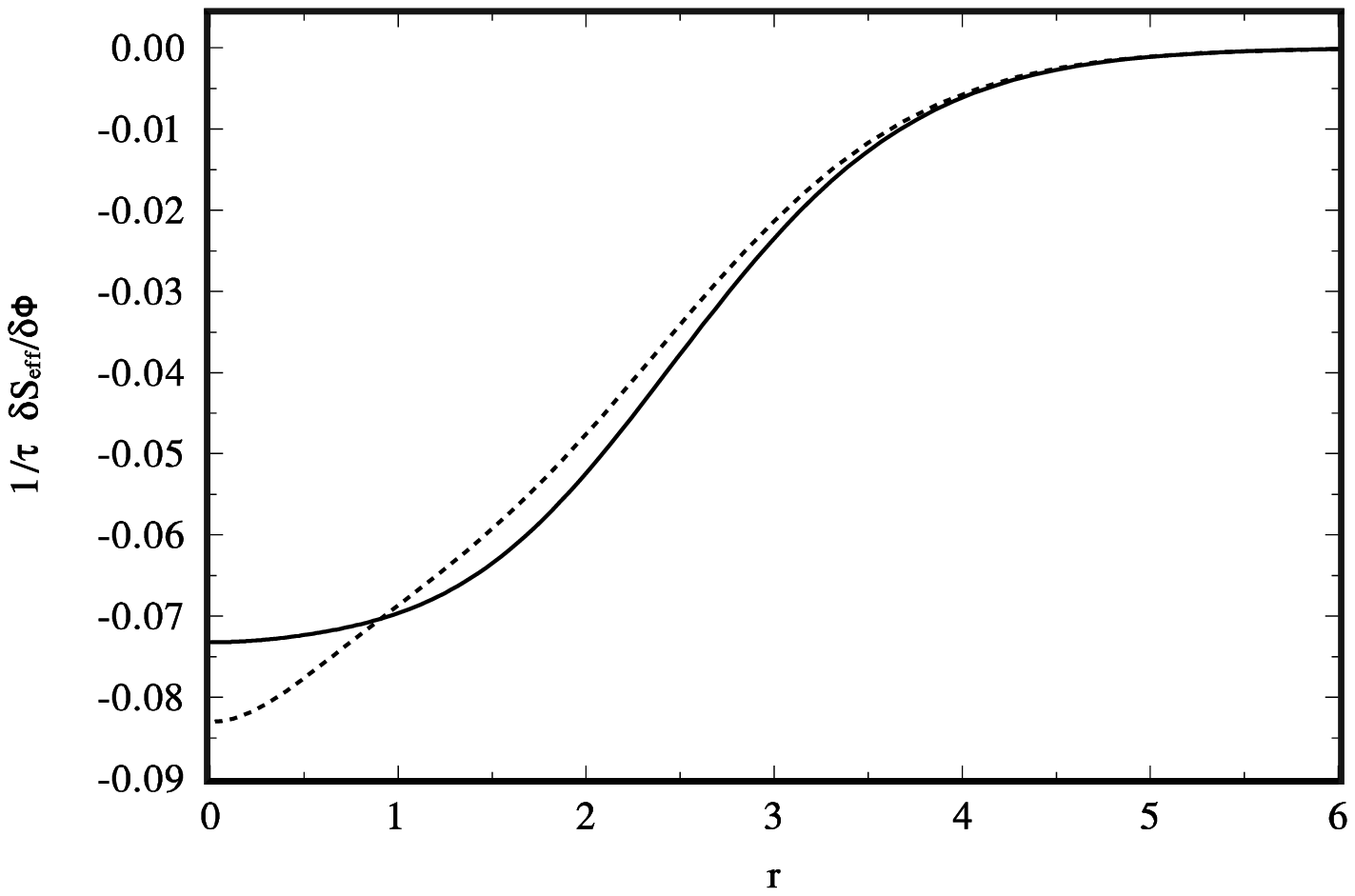}}
}

\vspace{3mm}

\centerline{\Large Figure 2}
\end{figure}
\clearpage
\begin{figure}
\centerline{
\mbox{\epsfxsize=16cm
\epsfbox{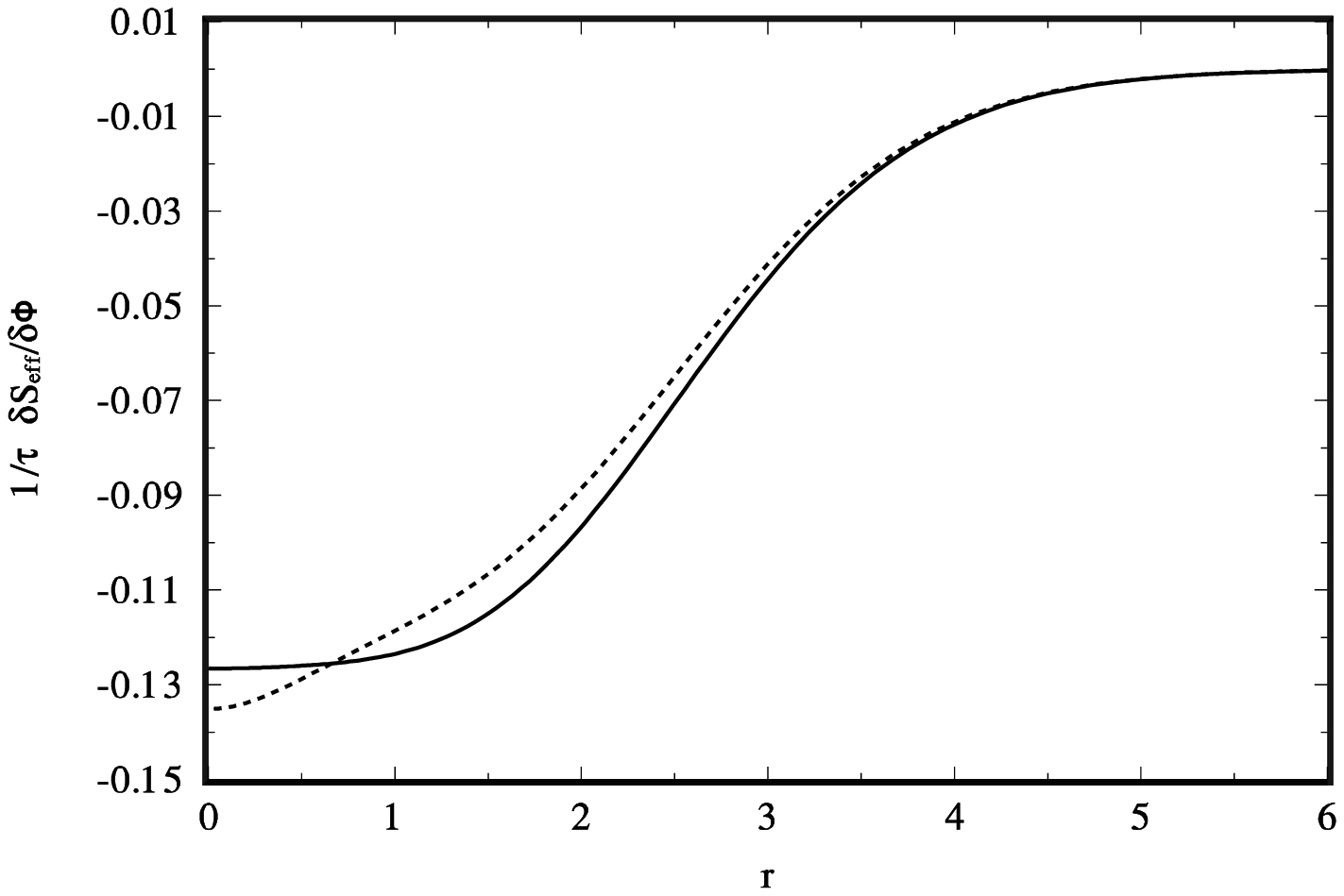}}
}

\vspace{3mm}

\centerline{\Large Figure 3}
\end{figure}
\clearpage

\begin{figure}
\centerline{
\mbox{\epsfxsize=16cm
\epsfbox{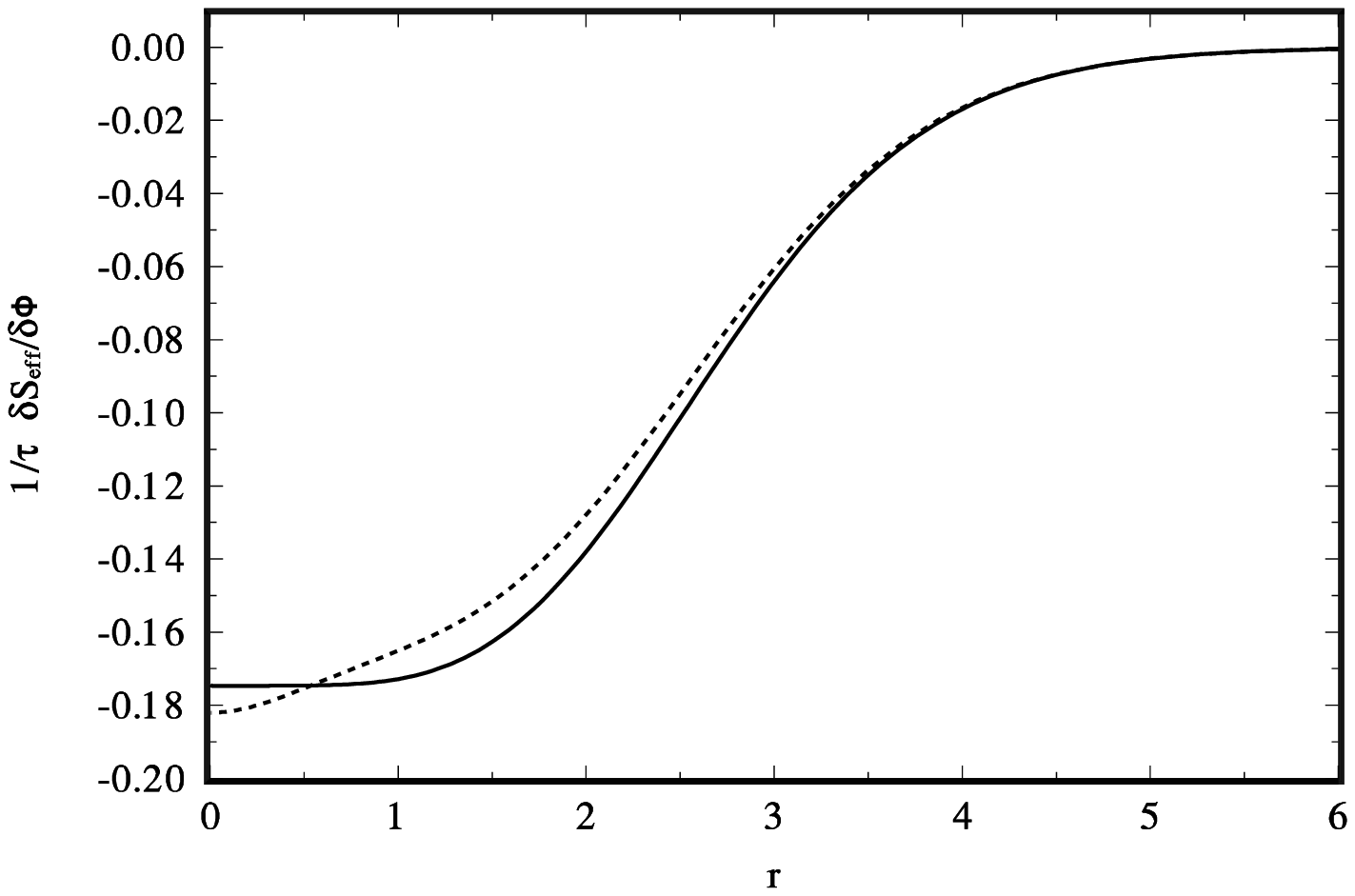}}
}

\vspace{3mm}

\centerline{\Large Figure 4}
\end{figure}

\clearpage
\begin{figure}
\centerline{
\mbox{\epsfxsize=16cm
\epsfbox{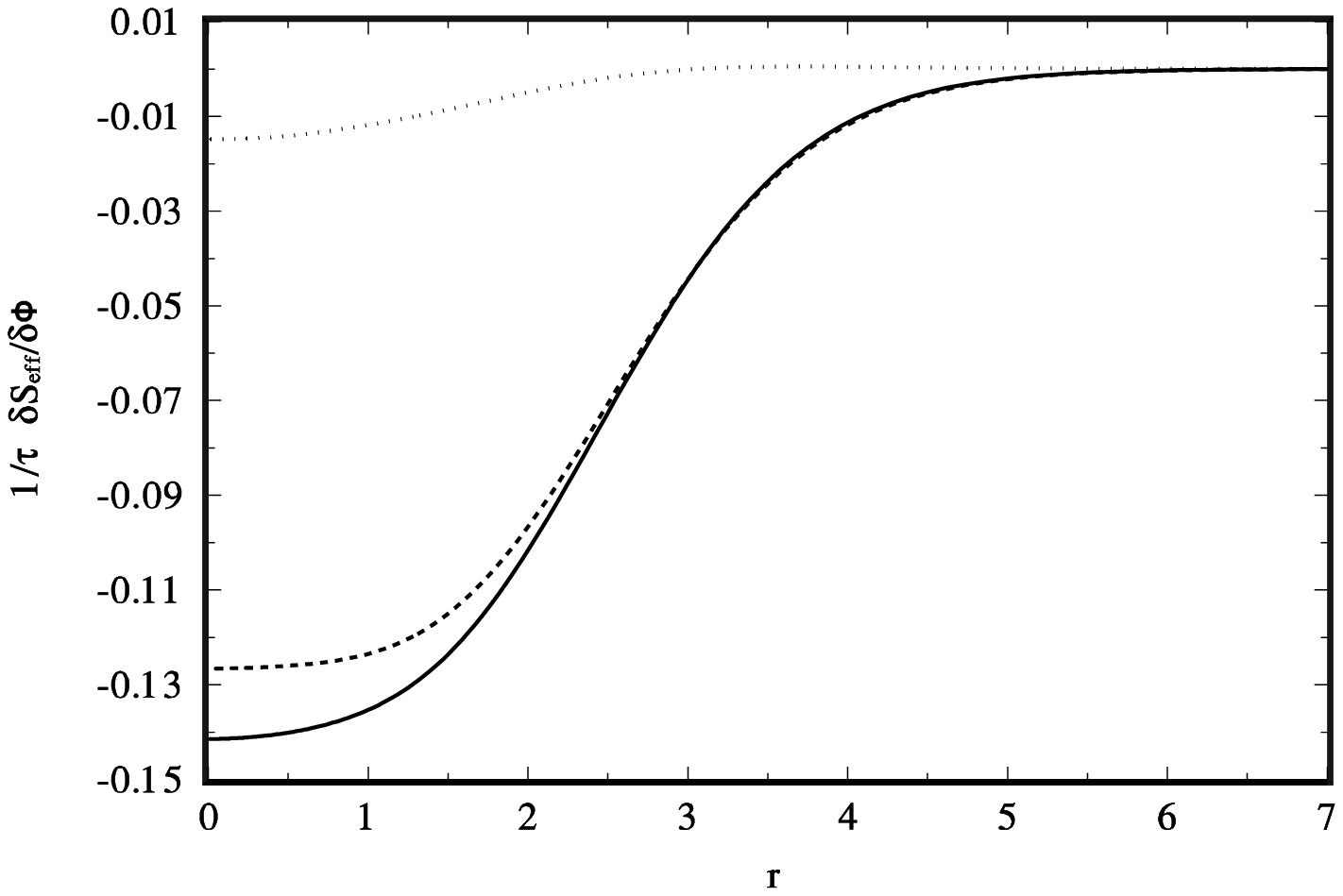}}
}

\vspace{3mm}

\centerline{\Large Figure 5}
\end{figure}
\end{document}